\documentclass{iaus}
\usepackage{graphicx}

\def\darcmin  {\hbox{$.\mkern-4mu^\prime$}}       
\def\kms {km s$^{-1}$}
\def\cm3 {cm$^{-3}$}
\def\cm2 {cm$^{-2}$}
\def \apj {ApJ}
\def \apjl {ApJL}
\def \aj {AJ}

\title[Extended SNR Masers] 
{Extended OH(1720 MHz) Maser Emission from Supernova Remnants}
\author[Hewitt et al.]   
{J.W. Hewitt$^1$, F. Yusef-Zadeh$^1$, M. Wardle$^2$  \and D.A. Roberts$^1$}
\affiliation{$^1$Department of Physics and Astronomy, Northwestern University, Evanston, IL 60608, USA \\[\affilskip]
$^2$Department of Physics, Macquarie University, Sydney, NSW 2109, Australia}

\pubyear{2007}
\volume{242}
\jname{Astrophysical Masers and their Environments IAU Symposium}
\editors{J.M. Chapman \& W.A. Baan, eds.}
\begin{document}

\maketitle

\begin{abstract}
Compact OH(1720 MHz) masers have proven to be excellent signposts for the interaction of supernova remnants with adjacent molecular clouds. 
Less appreciated has been the weak, extended OH(1720 MHz) emission which accompanies strong compact maser sources. Recent single-dish and interferometric observations reveal the majority of maser-emitting supernova remnants(SNRs) have accompanying regions of extended maser emission. Enhanced OH abundance created by the passing shock is observed both as maser emission and absorption against the strong background of the remnant. Modeling the observed OH profiles gives an estimate of the physical conditions in which weak, extended maser emission arises.
I will discuss how we can realize the utility of this extended maser emission, particularly the potential to measure the strength of the post-shock magnetic field via Zeeman splitting over these large-scales.
\keywords{supernova remnants, masers, ISM: magnetic fields}

\end{abstract}
\firstsection
\section{Introduction}

OH(1720 MHz) masers are an excellent indicator of a SNR shocking adjacent molecular material (see review by Brogan, elsewhere in these proceedings). The study of bright compact maser spots has successfully measured post-shock magnetic fields(e.g. Claussen et al 1997) and resolved the sizes and geometries of masing clumps in a number of remnants\cite{hoffmanphd}. Additionally, it has become increasingly apparent that weak, extended OH(1720 MHz) maser emission is also present on parsec scales in SNRs W28, G359.1-0.5, G357.7+0.3 and G357.7-0.1 (Yusef-Zadeh et al. 1999, 2003). This extended emission is well correlated with shocked molecular tracers (Lazendic et al 2002) and OH absorption in the other ground-state transitions. Here we discuss the frequency of extended maser emission in known maser-emitting SNRs as well as its utility in tracing the larger extent of the SNR/MC interaction and the potential to map the magnetic field via sensitive Zeeman measurements.

\section{Evidence of Extended OH(1720 MHz) Emission} \label{sec:eme}
To better determine the prevalence of extended OH(1720 MHz) emission we have used the 100m Green Bank Telescope(GBT) to map 15 maser-emitting remnants. At 1.7 GHz the GBT has a beamwidth of 7\darcmin2, so pointings are spaced by 3\darcmin3 to satisfy Nyquist sampling criterion and fully recover all flux from these sources. All four ground-state transitions of OH (1612, 1665, 1667, 1720 MHz) are observed simultaneously permitting OH absorption to be studied against the bright background of these remnants. Observations obtain a velocity resolution of 0.28\kms\ and a sensitivity of $\sim$20 mJy channel$^{-1}$ comparable to radio synthesis observations which first detected these masers (\cite[Frail et al 1994, Frail et al 1996, Green et al 1997, Koralesky et al 1998]{frail94,frail96,green97,koralesky98}).

It is clear that significant flux from 1720 MHz masers is missed by previous observations. A comparison of integrated maser line intensity between GBT and published VLA measurements is given in Table 1. We give the integrated line intensity\footnote{The integrated line intensity $\gamma = S_{p} \frac{2 (\Delta {\rm v})}{\sqrt{\pi/ln(2)}}$ is measured in units of Jy \kms\ where S$_{p}$ is the peak flux density and $\Delta$v is the full-width at half maximum as fit by a Gaussian. This is also commonly referred to as the $"$zeroeth moment$"$ of a Gaussian.} as it is a robust measure of maser emission less sensitive to the variable channel widths if the narrow maser line is not fully resolved. Absolute flux calibration for the GBT is estimated to be better than 10$\%$. Enhancements exceeding this uncertainty are interpreted as evidence for low gain, extended maser emission which is resolved out due to the interferometer's lack of sensitivity to large spatial scales and low brightness temperatures.

\begin{table}
\centering
\caption{Integrated Line Intensity of OH(1720 MHz) Emission}
\begin{tabular}{rrcccr}
\hline
SNR & Name & Velocity Range & GBT 0$^{th}$ Moment& VLA 0$^{th}$ Moment & Ref. \\
    &      & (\kms)     & (Jy \kms)  & (Jy \kms)     &  \\
\hline
6.4--0.1& W28 A,B  & +4.8,+6.3 & 15.31 & 7.826 & C97 \\ 
        &W28 C,D & +7.2,+14.1 & 11.57 & 6.980 & C97 \\ 
        & W28 E,F & +8.6,+15.9 & 273.2 & 176.0 & C97 \\ 
16.7+0.1 &         & +19.9 & 0.411 & 0.245 & G97\\
21.8--0.6 & Kes 69  & +69.8 & 0.130 & 0.093 & G97\\
 31.9+0.0 & 3C391 & +105.3 & 0.200 & 0.063 & F96\\
          &      & +109.8 & 0.255 & 0.077 & F96\\
 32.8--0.1 & Kes 78  & +86.2 & 0.523 & 0.510 & K98 \\
34.7--0.4 & W44 A  & +42.9       & 7.640 & 1.288 & C97 \\ 
          & W44 B,C & +44.7,+46.7 & 17.91 & 2.314 & C97 \\ 
          & W44 D,E,F  & +43.7,+46.9 & 48.91 & 33.41 & C97 \\ 
 49.2--0.7 & W51C    & +67.4,+74.3 & 7.815 & 8.308 & B00 \\
189.1+3.0 & IC 443 G& --4.6,0.5 & 3.916 & 3.168 & C97\\
          & IC 443 D& --9.9,--4.9 & 0.291 & -- & --\\ 
          & IC 443 B& --12.6,--4.9 & 0.223 & -- & --\\ 
348.5+0.1 & CTB 37A & --21.4 & 0.333 & 0.383 & B00 \\
          &         & --63.5,--66.2 & 1.671 & 1.702 & B00 \\
349.7+0.2 &         & +14.3,+16.7 & 3.528 & 2.153 & B00 \\
357.7--0.1 & Tornado & --12.8,--14.7 & 3.011 & 0.590 & F99 \\
357.7+0.3 & Square  & --34.1,--37.2 & 9.426 & 1.117 & F99 \\
\hline
\end{tabular}
\label{tbl:1720fits}
\\
{\small References: F96=\cite{frail96}; C97=\cite{claussen97}; G97=\cite{green97}; K98=\cite{koralesky98}; F99=\cite{fyz99}; B00=\cite{brogan00}.}
\end{table}

It is unlikely that the observed flux enhancement is a result of time-variable masing that has brightened significantly since VLA observations were conducted. \cite{goss68} first detected OH (1720 MHz) maser emission in W28 and no time variability has yet been reported for that source or any other(Frail et al. 1994). 

Thus extended OH(1720 MHz) maser emission appears to be a common phenomenon, detected in nine of twelve remnants considered here\footnote{We note that for three SNRs located near the Galactic Center (Sgr A East, Sgr D SNR and G1.4-0.1) absorption at 1720 MHz by intervening gas makes our measurements of maser intensity uncertain, and will be considered in a future work.}. Even more striking is our detection of new extended maser components in IC 443 that appear without associated compact maser spots\cite[(Hewitt et al. 2006)]{hewitt06}. Previously only masers associated with shocked molecular clump G had been detected. Narrow maser emission associated with clumps B and D is observed with the GBT yet no compact maser spots was detected toward these clumps down to 5 mJy beam$^{-1}$ channel$^{-1}$ with high-resolution VLA observations. These new masers are observed at a brightness level some 20 times lower than the bright masers in clump G but with narrow line widths($\sim$ 2 \kms ) and low brightness temperatures(T$_B$ $\le$ 2500 K) consistent with the properties of extend masers. Modeling discussed in the next section supports the idea that in clump G conditions are sufficient to obtain strong inversion of the 1720 MHz line, while in clumps B and D conditions are less ideal. 

\section{OH Modeling and Abundance} \label{sec:modeling}

OH pumping models indicate that the 1720 MHz line can be weakly inverted over a broad range of conditions provided the FIR continuum is not strong. Theoretical considerations give strict constraints on the conditions necessary for SNR OH (1720 MHz) masers: moderate temperatures of 50Ð-125 K, densities of 10$^5$ cm$^{-3}$, and OH column densities of the order 10$^{16}$ to 10$^{17}$ \cm2 (Lockett et al. 1999). Observations of all four OH ground-state transitions permits detailed modeling of the post-shock gas in which masers arise. Here we discuss modeling of masers detected toward IC 443 to illustrate the conditions for weak, extended OH(1720 MHz) emission.

\begin{figure}
\centering
(A)\ \ \ \ \ \ \ \ \ \ \ \ \ \ \ \ \ \ \ \ \ \ \ \ \ \ \ \ \ \ \ \ \ \ \ \ 
(B)\ \ \ \ \ \ \ \ \ \ \ \ \ \ \ \ \ \ \ \ \ \ \ \ \ \ \ \ \ \ \ \ \ \ \ \ 
(C)\ \ \ \ \ \ \ \ \ \ \ \ \ \ \ \ \ \ \ \ \ \ \ \ \ \ \ \ \ \ \ \ \ \ \ 
\\
\includegraphics[scale=0.28]{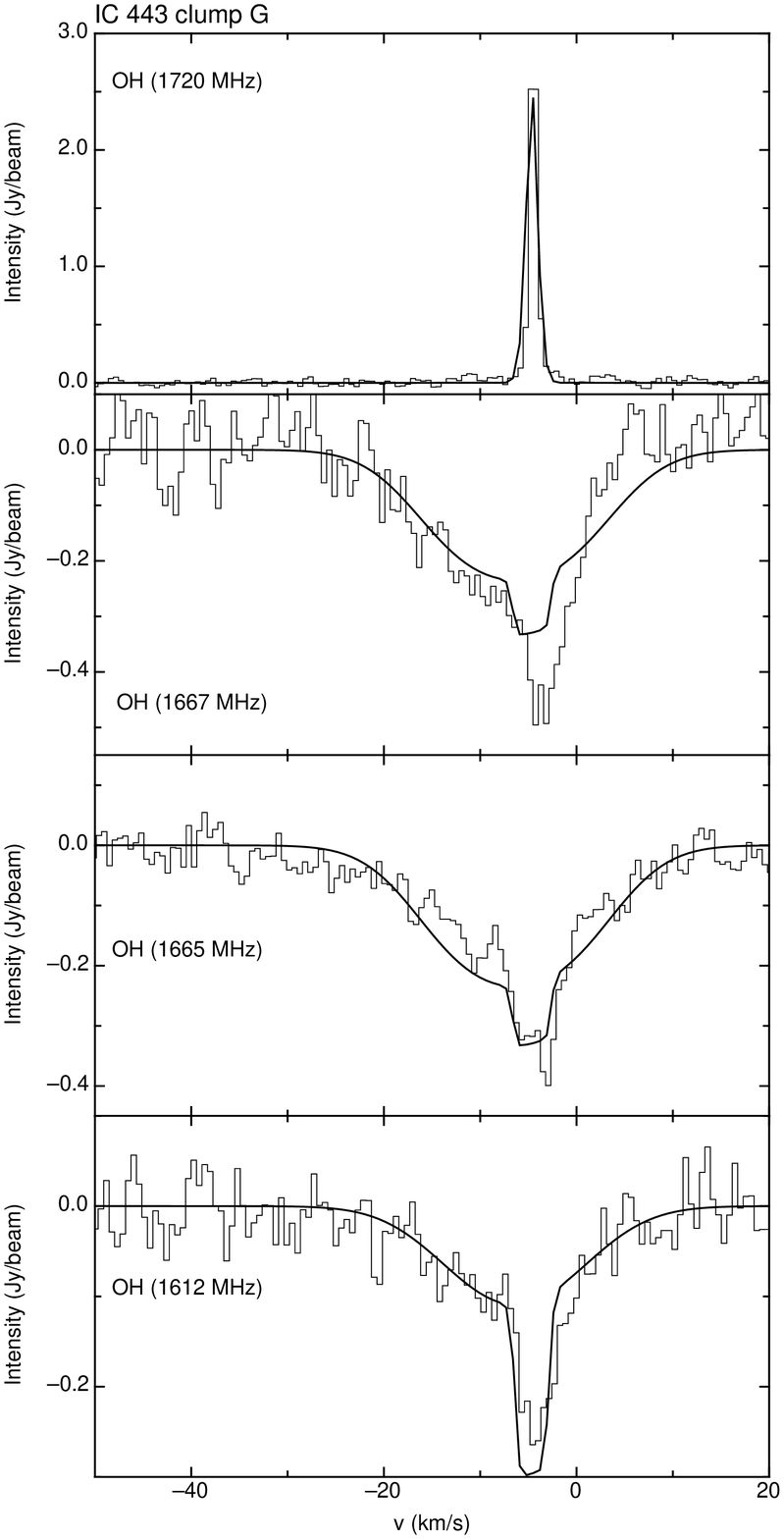}
\includegraphics[scale=0.28]{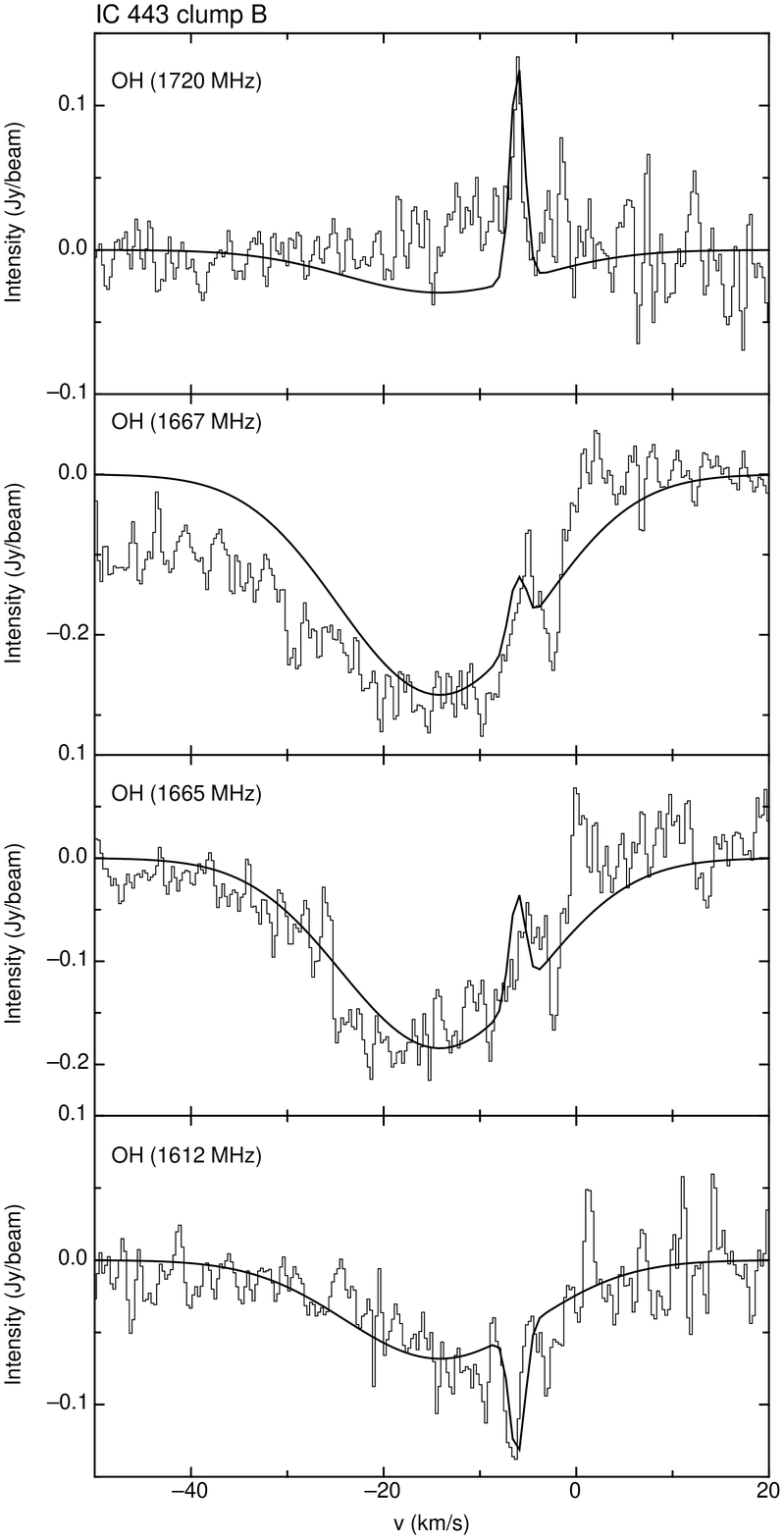}
\includegraphics[scale=0.28]{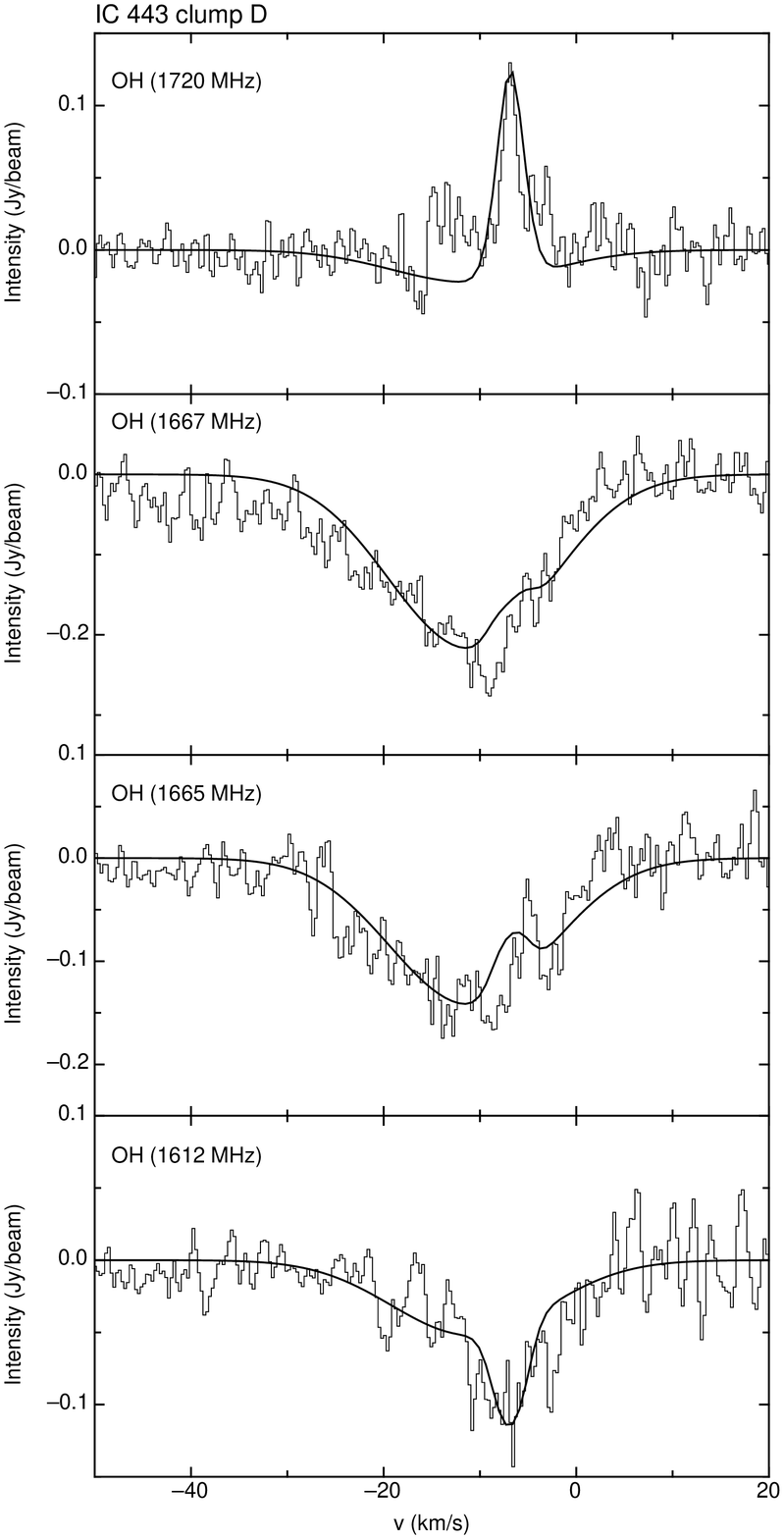}
\caption{Spectra of all four ground-state transitions of OH from pointings taken toward three clumps with detected 1720 MHz emission. The solid line shows the best-fit model for the observed spectrum. {\bf [A]} Clump G shows a strong, narrow emission feature is seen at 1720 MHz, whereas broad absorption is seen at the other transitions of OH, indicating a transverse shock. {\bf [B]} Clump B shows a narrow maser feature at 1720 MHz with broad, blue-shiftted absorption. {\bf [C]} Clump D also shows weak maser emission and broad absorption. }\label{fig:modelD}
\end{figure}

A simple two-component fit is sufficient to model line profiles of shocked gas towards both compact (G) and extended (B and D) masers in IC 443. As shown in figure~\ref{fig:modelD} we include components fit to both narrow maser emission and broad absorption arising from the shocked gas. Each component is characterized by its line-of-sight OH column density N$_{OH}$, kinetic temperature T$_{k}$, beam filling factor $f$, and mean and FWHM velocities, v$_0$ and $\Delta$v. For further details of the modeling procedure see \cite{hewitt06}.

The geometry of the shock can be roughly inferred from absorption profiles. For clump G the absorption is largely symmetric about the maser velocity indicating the shock is propagating transversely across the plane of the sky. Toward weak maser features in clumps B and D blue-shifted absorption features extending out to large negative velocities clearly indicate the shock is propagating with a significant inclination toward the observer's line of sight. Maser features appear only at velocities systemic to the remnant which is likely due to sufficient velocity coherence only being obtained in sufficiently transversely shocked clumps. Weak maser emission is seen extending along the southern shocked ridge of IC443 in figure 2. The observed line profiles for clumps B and D are best fit by T$_k$ = 35 K, N$_{OH}\approx$10$^{16}$ \cm2 \ and a low filling factor of order 10$\%$.
A comparable column is determined in the broad, blue-shifted absorption component but with a filling factor near unity. Bright maser clump G is fit by T$_k$ = 60 K and N$_{OH}\approx$10$^{17}$ \cm2 , physical conditions which yield a stronger 1720 MHz line inversion. In IC 443 it appears that weak maser emission results from physical conditions and a shock geometry which are less optimal than for bright compact masers.

To obtain the high OH columns observed it had been expected that C-type shock waves would very efficiently convert the bulk of gas-phase oxygen into water, a few percent of which could be dissociated into OH just behind the shock\cite[(Wardle 1999)]{wardle99}. Interestingly, SWAS observations of shocked water give columns of only a few$\times$10$^{13}$ cm$^{-2}$,  yielding a post-shock water abundance of between 10$^{-8}$ and 10$^{-7}$\cite[(Snell et al 2005)]{snell05}. This is orders of magnitude lower than the post-shock abundance of OH of a few $\times$10$^{-6}$ inferred from our observations\cite[(Hewitt et al 2006)]{hewitt06}. This apparent lack of water is not accounted for in any present theory for the production of OH (1720 MHz) masers. As the oxygen chemistry is well known, resolving this riddle of missing water likely lies in a better understanding of the kinds of shocks present and the freeze-out of water ice onto grain surfaces.

\section{Zeeman Mapping of Extended Maser Emission} \label{sec:zeemap}
\begin{figure}
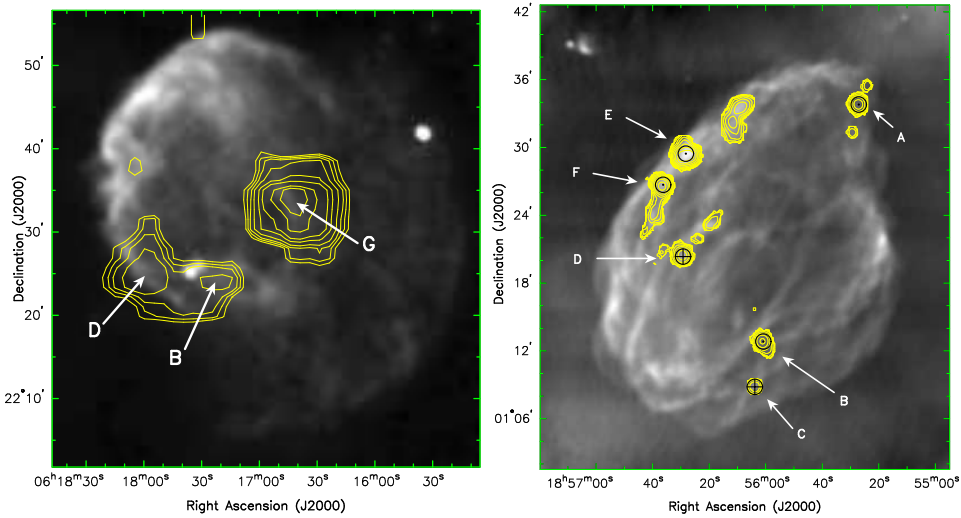

\centering
\includegraphics[scale=0.35]{iau242fig2a.ps}
\includegraphics[scale=0.335]{fig_w44.ps}
\caption{{\bf [Left]} Greyscale image of 327 MHz continuum from IC 443. Contours of the integrated intensity of OH(1720 MHz) emission detected by the GBT are shown with maser features B, D and G identified. {\bf [Right]} Greyscale of 1.4 GHz continuum from W44 \cite[(Giancani et al 1997)]{giacani97}. Contours of integrated maser intensity detected with preliminary VLA observations are superposed with maser features A-F identified. Here the line of sight magnetic field determined for each maser is denoted by symbols ($\odot$ = toward observer [negative field] and $\oplus$ = away from observer [positive field]).}\label{fig:w44}
\end{figure}

As the distribution of weak, extended maser emission appears to be a widespread tracer of the large-scale shock distribution, Zeeman splitting measurements of this emission allow us to probe the structure of the magnetic field behind the shock. As an example, results of sensitive Zeeman measurements of W44 with the VLA is given in figure~\ref{fig:w44}(Yusef-Zadeh, Roberts \& Hewitt 2007,  in
preparation). Here the line of sight magnetic field measured from fitting the Zeeman effect on the stokes V signal gives field strengths from --828 to +864 $\mu$G. The direction of the field with respect to the observer is seen to reverse direction on scales of a few parsecs. Extended OH(1720 MHz) emission is clearly seen as an arc connecting compact maser spots D, E and F. Near-IR H$_2$ and Spitzer 4.5$\mu$m images show shocked filaments corresponding to the observed extended maser emission\cite[(Reach et al 2006)]{reach06}. Given W44's relatively nearby location, strong continuum and large angular size it is particularly well suited for sensitive Zeeman measurements of the weak, extended maser emission. Typically observations of OH(1720 MHz) masers are done with snapshots or highly spaced configurations that are not sensitive to such extended structures. Sensitive Zeeman measurements of the extended maser component are an exciting direction of future study that hold the promise of resolving the structure of the magnetic field which dominates SNR shock interactions.

\begin{acknowledgments}
We thank Ron Maddalena, Jim Braatz and the GBT operators for their help throughout these observations. Support for this work was provided by the NSF through award GSSP06-0009 from the NRAO. JWH acknowledges partial support form an IAU Grant.
\end{acknowledgments}

\end{document}